\documentclass[10pt,preprint]{emulateapj}  

\newcommand{\ud}{{\rm d}} 

\newcommand{\bmath}[1]{\mbox{\boldmath{$#1$}}}
\newcommand{\bn}{\bmath{n}}
\newcommand{\nx}{n_{\rm x}}
\newcommand{\ny}{n_{\rm y}}

\def\comp{\,c/\omega_{\rm p}}
\def\ompt{\omega_{\rm p}t}

\newcommand{\gb}[1]{\gamma\beta_{\rm {#1}}}

\newcommand{\eq}[1]{eq.~(\ref{eq:#1})}
\newcommand{\fig}[1]{Fig.~\ref{fig:#1}}
\newcommand{\ditto}[1]{{_{\rm{#1}}}} 

\begin{document}
\title{Synthetic Spectra from PIC Simulations of Relativistic Collisionless Shocks}
\author{Lorenzo Sironi and Anatoly Spitkovsky}
\affil{Department of Astrophysical Sciences, Princeton University, Princeton, NJ 08544-1001}
\email{lsironi@astro.princeton.edu; \\ anatoly@astro.princeton.edu}

\begin{abstract}
We extract synthetic photon spectra from first-principles particle-in-cell simulations of relativistic  shocks propagating in unmagnetized pair plasmas. The two basic ingredients for the radiation, namely accelerated particles and magnetic fields, are produced self-consistently as part of the shock evolution. We use the method of \citet{hededal_05} and compute the photon spectrum via Fourier transform of the electric far-field from a large number of particles, sampled directly from the simulation. We find that the spectrum from relativistic collisionless shocks is entirely consistent with synchrotron radiation in the magnetic fields generated by Weibel instability. We can recover the so-called ``jitter'' regime only if we artificially reduce the strength of the electromagnetic fields, such that the wiggler parameter $K\equiv qB\lambda/mc^2$ becomes much smaller than unity ($B$ and $\lambda$ are the strength and scale of the magnetic turbulence, respectively). These findings may place constraints on the origin of non-thermal emission in astrophysics, especially for the interpretation of the hard (harder than synchrotron) low-frequency spectrum of Gamma-Ray Bursts.
\end{abstract}


\section{Introduction}\label{sec:intro}
Non-thermal photon spectra from Pulsar Wind Nebulae, jets from Active Galactic Nuclei, Gamma-Ray Bursts and Supernova Remnants are usually explained as synchrotron radiation from a power-law population of particles, presumably accelerated in collisionless shocks. The microphysical details of shock acceleration are still poorly known, however, and are the subject of active research. 

Particle-in-cell (PIC) simulations of colliding plasma shells have shown that Weibel instability \citep{weibel_59, medvedev_loeb_99, gruzinov_waxman_99} converts the free energy of counter-streaming flows into small scale (skin-depth) magnetic fields \citep{nishikawa_03, nishikawa_05, silva_03, frederiksen_04, hededal_04}. The  fields grow to sub-equipartition levels and deflect and randomize the bulk  flow, thus creating a shock \citep{spitkovsky_05, spitkovsky_08, chang_08, keshet_09}. A few percent of the incoming particles repeatedly scatter off the magnetic turbulence created by Weibel instability, and eventually populate a power-law high-energy tail in the particle spectrum behind the shock \citep{spitkovsky_08b,martins_09,sironi_spitkovsky_09}. 

Since most of the magnetic power generated by Weibel instability is concentrated on scales as small as a few plasma skin depths, it has been speculated that the emission mechanism in unmagnetized collisionless shocks may be the so-called ``jitter'' radiation. Whereas the standard synchrotron emission applies to large-scale fields, the jitter regime is realized if the scale $\lambda$ of the turbulence is such that the wiggler parameter $K\equiv qB\lambda/mc^2\ll1$ \citep{medvedev_00, medvedev_06, fleishman_06b, fleishman_06a}. Jitter radiation has been proposed as a solution for the so-called ``line of death'' puzzle in Gamma-Ray Burst (GRB) spectra, which below the peak frequency are sometimes harder than expected from synchrotron radiation \citep[e.g.,][]{preece_98}. Since PIC simulations self-consistently provide both the strength and the spatial structure of electromagnetic fields, as well as the particle distribution, it is possible to calculate the photon spectrum from first principles, thus determining whether the emission is more synchrotron-like or jitter-like. 

In this work, we present synthetic spectra extracted from PIC simulations of relativistic unmagnetized collisionless shocks. 
In \S 2 we show the simulated shock structure and particle energy spectrum. The numerical technique that we employ to compute the photon spectrum is described in \S 3.1, and in \S 3.2 we show our results. 
In \S4 we discuss the implications of our findings for the interpretation of the non-thermal radiation from astrophysical sources.

\begin{figure}[tbp]
\begin{center}
\includegraphics[width=0.5\textwidth]{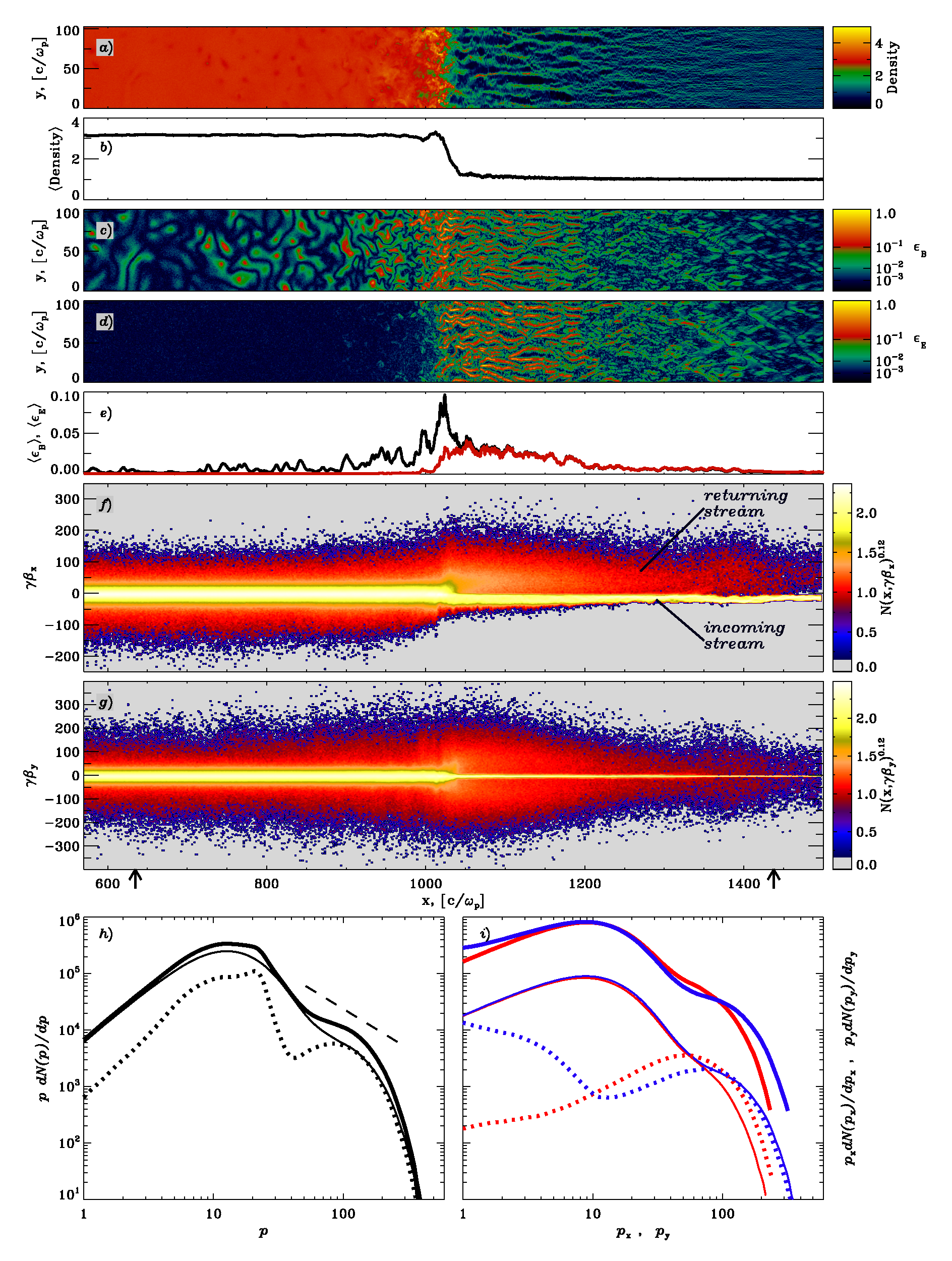}
\caption{Shock structure and particle energy spectra at time $\ompt=2250$. a) Number density in the simulation plane, normalized to the upstream density. b) Transversely-averaged density. c)-d) Magnetic and electric energy density in the simulation plane, normalized to the upstream kinetic energy density. e) Transversely-averaged magnetic (black) and electric (red) energy density. f) Longitudinal phase space for positrons, shown as a 2D histogram (electron phase space is nearly identical). g) Transverse phase space for positrons. h) Positron spectrum $p\,\ud N/\ud p$ versus $p$ (thick solid line; $p\equiv\gamma\beta=[(\gb{x})^2+(\gb{y})^2]^{1/2}$ is the particle 4-velocity) in the region $x_{\rm sh}-400\comp<x<x_{\rm sh}+400\comp$ (delimited by the arrows at the bottom of panel (g)), and the relative contributions of particles behind (thin solid line) and ahead (dotted line) of the shock. The dashed line corresponds to a power-law distribution $\ud N/\ud p\propto p^{-\alpha}$ with $\alpha=2.5$. i) Positron spectrum $p_{\rm x}\,\ud N/\ud p_{\rm x}$ versus $p_{\rm x}$ (with $p_{\rm x}\equiv\gb{x}$) for particles with $p_{\rm x}>0$ (red lines), and positron spectrum $p_{\rm y}\,\ud N/\ud p_{\rm y}$ versus $p_{\rm y}$ (with $p_{\rm y}\equiv\gb{y}$) for particles with $p_{\rm y}>0$ (blue lines). The line style coding is the same as in panel (h). The total spectrum around the shock  (thick solid lines) is shifted upward by a factor of 10 for clarity.
}
\label{fig:fluid}
\end{center}
\end{figure}

\section{Shock Structure and Particle Energy Spectrum}
We use the three-dimensional (3D) electromagnetic PIC code TRISTAN-MP \citep{buneman_93, spitkovsky_05} to simulate a relativistic shock propagating into an unmagnetized pair plasma. The shock is triggered by reflecting an incoming cold ``upstream''  flow off a conducting wall at $x = 0$ \citep[e.g.,][]{sironi_spitkovsky_09}. The simulation is performed in the ``wall'' or ``downstream'' frame. The incoming flow propagates along $-\bmath{\hat{x}}$ with Lorentz factor $\gamma_0 = 15$, and the shock moves along $+\bmath{\hat{x}}$.

To follow the shock evolution for longer times with fixed computational resources, we use a 2D simulation domain in the $xy$ plane. 
In the case of an unmagnetized 2D shock, only the in-plane components of the velocity, current and electric field, and only the out-of-plane component of the magnetic field are present. Each computational cell is initialized with 16 particles per species. The relativistic plasma skin depth for the upstream flow ($c/\omega_{\rm p}$)  is resolved with 10 cells and the simulation timestep is $\Delta t=0.045\,\omega_{\rm p}^{-1}$. The simulation box is $100\,c/\omega_{\rm p}$ wide (along $y$) and, at the final simulation time $\omega_{\rm p}t=4500$, it is $4500\comp$ long (along $x$).

In \fig{fluid} we show the internal structure of the shock at time $\ompt=2250$. The 2D plots of number density (\fig{fluid}a), magnetic energy (\fig{fluid}c) and electric energy (\fig{fluid}d) show the filaments produced by Weibel instability  in the upstream region, with characteristic transverse scale $\sim10\comp$. The magnetic filaments are advected with the upstream flow, so in the  simulation frame their magnetic and electric components are  comparable (black and red line in \fig{fluid}e, respectively). At the shock ($x_{\rm sh}\simeq1030\comp$ at time $\ompt=2250$), the filaments merge and the magnetic energy density peaks at $\sim10\%$ of the upstream kinetic energy density (\fig{fluid}e). The magnetic field decays farther downstream, where the field is confined within islands of typical scale $\sim20\comp$ (\fig{fluid}c). 

The particle energy spectrum behind the shock ($x_{\rm sh}-400\comp<x<x_{\rm sh}$; thin solid line in \fig{fluid}h) consists of a relativistic Maxwellian and a high-energy tail, which can be fitted as a power-law of index $\alpha=2.5$ (dashed line in \fig{fluid}h) with an exponential cutoff \citep{spitkovsky_08b}. In the upstream spectrum ($x_{\rm sh}<x<x_{\rm sh}+400\comp$; dotted line in \fig{fluid}h), the unshocked beam populates the low-energy peak at $p\equiv\gamma\beta\simeq15$, whereas the shock-accelerated returning particles with $\gb{x}>0$ (see the hot, diffuse  population in \fig{fluid}f) appear as as a high-energy bump. Since at the highest energies the particles ahead of the shock account for nearly half of the total census (compare dotted and thin solid lines in \fig{fluid}h), the high-energy tail in the total spectrum ($x_{\rm sh}-400\comp<x<x_{\rm sh}+400\comp$; thick solid line in \fig{fluid}h) is significantly flatter than in the downstream spectrum.

\section{Synthetic Photon Spectrum}
\subsection{Numerical Technique}\label{sec:method}
We summarize the method introduced by \citet{hededal_phd} and \citet{ hededal_05} \citep[see also][]{nishikawa_09,martins_09b} to extract synthetic spectra from simulations of collisionless shocks. The electric far-field from a particle with charge $q$, velocity $\bmath{v}=\bmath{\beta}c$ and acceleration $\dot{\bmath{v}}=\dot{\bmath{\beta}}c$ is \citep{jackson_99}
\begin{equation}\label{eq:efield}
\bmath{E}(\bmath{x},t)=\frac{q}{c}\left[\frac{\bmath{n}\times\{(\bmath{n}-\bmath{\beta})\times\dot{\bmath{\beta}}\}}{(1-\bmath{\beta}\cdot\bmath{n})^3R}\right]_{\rm{ret}}~,
\end{equation}
where the unit vector $\bmath{n}$ points toward the observer, at distance $R$ from the emitting particle. Here, the quantity in square brackets is to be evaluated at the retarded time $t'=t-R(t')/c$.  
The photon spectrum is then computed via the Fourier transform of Poynting flux associated with the field in eq.~(\ref{eq:efield}). The energy $\ud W$ received per unit solid angle $\ud \Omega$ (around the direction $\bmath{n}$) and per unit frequency $\ud \omega$ can be computed as \citep{jackson_99}
 \begin{equation}\label{eq:rad}
\frac{\ud^2 W}{\ud \Omega \ud \omega}\!\!=\!\!\frac{q^2}{4\pi^2 c}\!\left|\int_{_{-\infty}}^{^{+\infty}}\!\!\!\!\!\!\frac{\bmath{n}\times\{(\bmath{n}-\bmath{\beta})\times\dot{\bmath{\beta}}\}}{(1-\bmath{\beta}\cdot\bmath{n})^2}e^{i\omega(t'-\scriptsize{\bmath{n}\cdot\bmath{r}(t')/c})}\ud t'\right|^2
\end{equation}
where $\bmath{r}(t')$ is the particle trajectory. Here, we neglected the Tsytovich-Razin effect due to the dispersive properties of the plasma \citep{rybicki_lightman_79}.

In PIC simulations we know the positions, velocities and accelerations of simulation particles with time resolution $\Delta t=0.045\,\omega_{\rm p}^{-1}$. In order to accurately compute the integral in \eq{rad}, we interpolate the orbit of the selected particles so that to achieve an effective timestep of $0.1\,\Delta t$. For a given choice of $\bmath{n}$, we can then integrate eq.~(\ref{eq:rad})  to obtain the photon spectrum from each particle. Assuming that the  far-fields by different particles are phase-uncorrelated, the total spectrum will be the sum of the spectra of individual particles.

We have implemented eq.~(\ref{eq:rad}) and tested it for the cases of synchrotron, bremsstrahlung and wiggler/undulator radiation, finding good agreement with analytic solutions. Following  \citet{hededal_phd} and \citet{ hededal_05}, we have traced test particle orbits in magnetic turbulence with a \textit{prescribed} wave spectrum, and we have verified the transition from the synchrotron to the jitter regime as the wiggler parameter $K$ becomes smaller than unity.  

We now present the photon spectrum resulting from $\sim10,000$ particles moving in the electromagnetic fields \textit{self-consistently} produced in our PIC simulations. We typically follow the particle trajectories over $3000\,\Delta t=135\,\omega_{\rm p}^{-1}$, which is long enough to reach convergence in the shape of the spectrum, but short compared to the characteristic  time of the shock evolution, such that the spectrum we obtain may be regarded as ``instantaneous.'' We compute the spectrum for both head-on emission ($\bn=\bmath{\hat{x}}$, which we call ``$\nx=1$'' from now on) and edge-on emission ($\bn=\bmath{\hat{y}}$, ``$\ny=1$'' from now on). Our calculations are performed in the downstream fluid frame. An additional Lorentz transformation is required if the downstream medium is moving with respect to the observer.

\subsection{Results}\label{sec:results}

\begin{figure}[tbp]
\begin{center}
\includegraphics[width=0.5\textwidth]{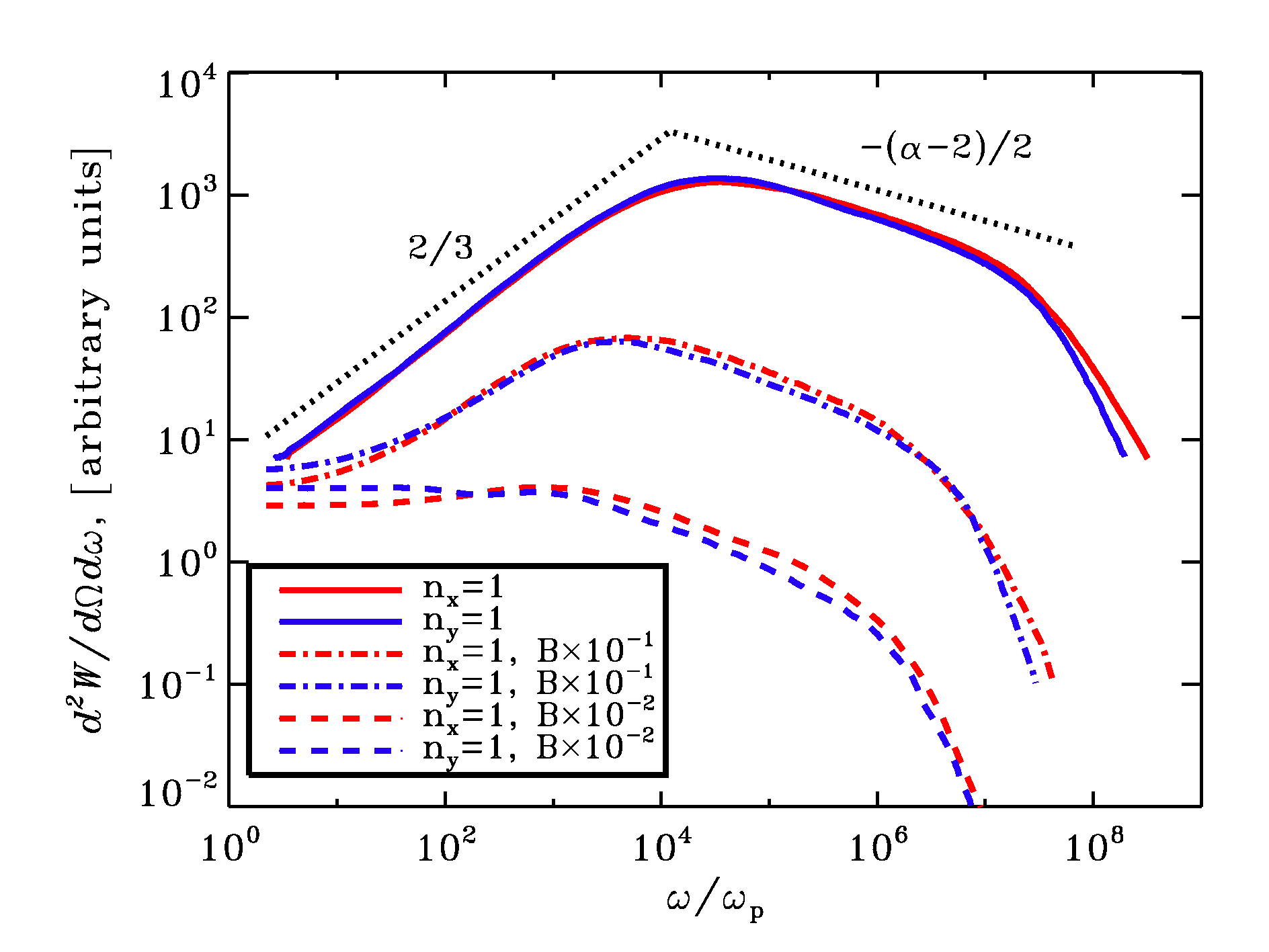}
\caption{Solid lines: photon spectrum from a power-law population of particles ($\ud N/\ud p\propto p^{-\alpha}$ between $p_{\rm min}=50$ and $p_{\rm max}=2000$, with $\alpha=2.5$) injected behind the shock ($x_{\rm sh}-400\comp<x<x_{\rm sh}$). The particles are traced in the fields of the simulation at time $\ompt=2250$. The photon spectrum from the same particle population, but evolved in electromagnetic fields artificially reduced by a factor of 10 (100, respectively), is shown as dot-dashed (dashed, respectively) lines. Red lines are for head-on emission ($\bn=\bmath{\hat{x}}$), blue lines for edge-on emission ($\bn=\bmath{\hat{y}}$).}
\label{fig:powlaw}
\end{center}
\end{figure}

We have studied the emission from relativistic shocks with two experiments. First, we have injected a power-law distribution of particles into the downstream region, and we have traced their orbits in a fixed snapshot of the fields of the simulation. The photon spectrum is then calculated with the technique described above.
Second, we compute the spectrum from trajectories of particles extracted directly from the simulation. In this case, both the particle distribution and the electromagnetic turbulence are provided by the PIC simulation, and the resulting emission will be the \textit{self-consistent} spectrum from our relativistic shock. 

\fig{powlaw} shows the emission spectrum from a power-law population of particles behind the shock ($x_{\rm sh}-400\comp<x<x_{\rm sh}$). We inject a 2D isotropic distribution with $\ud N/\ud p\propto p^{-\alpha}$ ($\alpha=2.5$), between $p_{\rm min}=50$ and $p_{\rm max}=2000$. The spectral index and the lower cutoff of the distribution are chosen to mimic the high-energy tail of the downstream particle spectrum in the simulation (thin solid line in \fig{fluid}h). The injected particles are evolved in a fixed snapshot of the electromagnetic fields from the PIC simulation, at time $\ompt=2250$. 

Since electric fields are negligible in the downstream medium (see \fig{fluid}d-e), the resulting photon spectrum will probe the strength and structure of the magnetic fields, and it will clarify which regime -- synchrotron or jitter -- is appropriate to describe the particle emission. The solid lines in \fig{powlaw} show that the spectrum can be well approximated by two power-law segments. Regardless of the observer's direction $\bmath{n}$ (red line for $n_{\rm x}=1$, blue line for $n_{\rm y}=1$), the slope at the low frequencies is remarkably close to $2/3$ (dotted line in \fig{powlaw}), as expected for synchrotron emission from a 2D particle distribution \citep[e.g.,][]{jackson_99}.\footnote{For a 3D distribution the low-frequency spectrum is $\propto\omega^{1/3}$.} In $N$ dimensions, the high-frequency slope should be $-[\alpha-(4-N)]/2$, which reduces to $-(\alpha-2)/2=-0.25$ for $N=2$ and $\alpha=2.5$, in agreement with our spectra (see dotted line). The similarity between the cases $\nx=1$ and $\ny=1$ suggests, given the isotropy of the injected particle distribution,  that the downstream magnetic fluctuations are spatially isotropic, as seen in \fig{fluid}c.\footnote{The slight difference at high frequencies between $\nx=1$ and $\ny=1$ is due to the residual electric fields at $x\lesssim x_{\rm sh}$ (see \fig{fluid}d-e), and it disappears if electric fields are neglected  while computing the spectrum.}
  
A transition to the jitter regime should appear when the wiggler parameter $K$ becomes significantly smaller than unity \citep{medvedev_00, medvedev_06, fleishman_06b, fleishman_06a}. We have tried to artificially lower the value of $K$ by decreasing the strength of the electromagnetic fields, by a factor of 10 (dot-dashed lines in \fig{powlaw}) and 100 (dashed lines in \fig{powlaw}). The high-frequency power-law decreases in intensity and shifts to lower frequencies, proportionally to the average magnetic field. The low-frequency spectrum becomes softer for decreasing $B$, approaching the flat slope expected in the jitter regime when the shock is viewed edge-on. When we impose a magnetic field spectrum of the form $\bmath{B}(\bmath{k})\propto\delta(\hat{\bmath{k}}-\hat{\bmath{k}}_0)$ (here, $\hat{\bmath{k}}_0$ is a fixed direction in $k$-space) with wiggler parameter $K\ll1$, we are able to recover the hard low-frequency slope ($\propto\omega^1$) discussed by \citet{medvedev_00} for head-on emission from shocks. However, we do not observe it for the downstream turbulence self-consistently generated in the simulation, suggesting that the magnetic field fluctuations are not sufficiently ordered. 

\begin{figure}[tbp]
\begin{center}
\includegraphics[width=0.5\textwidth]{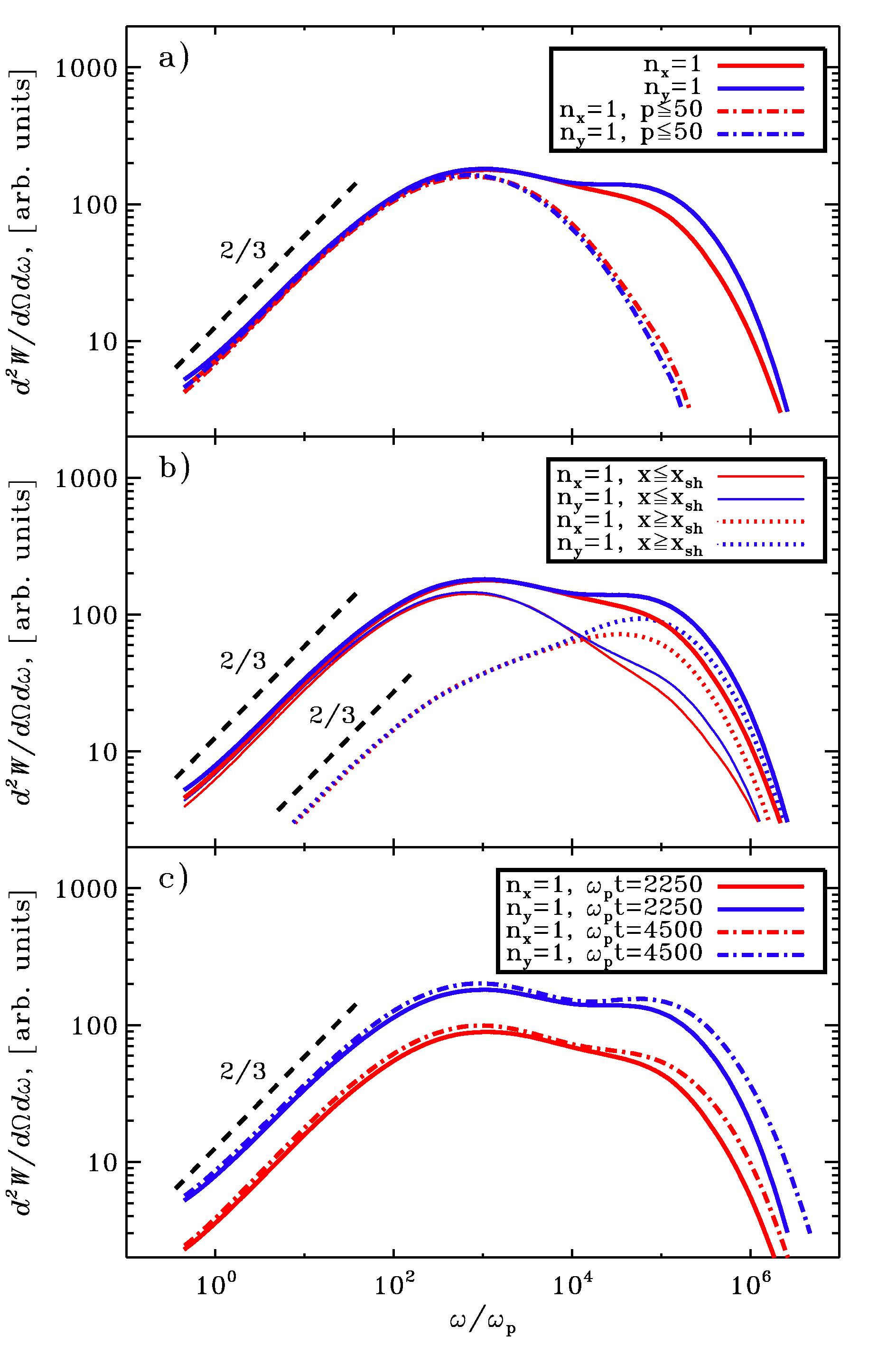}
\caption{Photon spectrum (thick solid lines, in all panels) for particles extracted from the PIC simulation, evolved in time-varying electromagnetic fields around $\ompt=2250$. Red lines are for head-on emission ($\bn=\bmath{\hat{x}}$), blue lines for edge-on emission ($\bn=\bmath{\hat{y}}$). a) Dot-dashed lines only include the contribution of thermal particles, with 4-velocity $p\leq50$. b) Relative contribution of downstream (thin solid lines) and upstream (dotted lines) particles. c) Time evolution of the photon spectrum, from $\ompt=2250$ (thick solid lines) to $\ompt=4500$ (dot-dashed lines). Here, the case $\nx=1$ has been shifted downward by a factor of 2 for clarity.}
\label{fig:simprt}
\end{center}
\end{figure}

In \fig{simprt} we show the photon spectrum resulting from a sample of particles extracted directly from the PIC simulation, followed near $\ompt=2250$ in the \textit{time-varying} electromagnetic fields of the simulation. The selected particles start in the region $x_{\rm sh}-400\comp<x<x_{\rm sh}+400$. The photon spectrum (thick solid lines in \fig{simprt}; red for $\nx=1$, blue for $\ny=1$) confirms  that the emission occurs in the synchrotron regime, as the $2/3$ slope at the low frequencies suggests (black dashed line). 

Most of the low-frequency emission is powered by the thermal particles behind the shock, with 4-velocity $p\leq50$ (dot-dashed lines in \fig{simprt}a). At  high frequencies, the emission along $\bmath{\hat{y}}$ (blue solid line in \fig{simprt}a) is more powerful than along $\bmath{\hat{x}}$ (red solid line in \fig{simprt}a). This reflects the fact that the highest energy particles are grazing the shock surface \citep{spitkovsky_08b} and they mostly contribute to the emission along $\bmath{\hat{y}}$. The difference between $\nx=1$ and $\ny=1$ in \fig{simprt}a can be directly related to the difference between the particle spectra $p_{\rm x}\,\ud N/\ud p_{\rm x}$ and $p_{\rm y}\,\ud N/\ud p_{\rm y}$ shown in \fig{fluid}i (thick solid lines, respectively red and blue). In fact,  due to relativistic beaming, an observer located along $\bmath{\hat{x}}$ will more likely detect the radiation from particles with $p_{\rm x}/p_{\rm y}\gg1$ (and $p_{\rm x}>0$), whereas  the particles with $p_{\rm y}/p_{\rm x}\gg1$ (and $p_{\rm y}>0$) will contribute more to the emission along $\bmath{\hat{y }}$. 

As shown in \fig{simprt}b, the spectrum of downstream particles (thin solid lines)  dominates the total emission (thick solid lines) in the low-frequency bump, whereas the high frequencies are mostly powered by the spectrum of upstream particles (dotted lines). This is due to a combination of two effects: first, at the highest particle energies the upstream particles may dominate by number (compare red dotted and thin solid lines in \fig{fluid}i); second, the electromagnetic energy in the upstream region is on average larger than in the downstream, since upstream electric fields are as strong as magnetic fields, and also because the field is built up ahead of the shock on a length scale larger than the decay scale in the downstream (\fig{fluid}e). It is worth pointing out that, at low frequencies,  the spectra of both downstream and upstream particles can be fitted with the synchrotron slope 2/3 (black dashed lines).

\fig{simprt}c shows the time evolution of the total photon spectrum, from $\ompt=2250$ (thick solid lines) to $\ompt=4500$ (dot-dashed lines). The upper cutoff in the particle energy spectrum grows linearly with time \citep{spitkovsky_08b, sironi_spitkovsky_09}, and as a result the photon spectrum extends to higher frequencies. Also, the spectral intensity increases, especially at high frequencies, due to stronger electromagnetic fields and a larger number of accelerated particles.

\section{Discussion}
We have computed synthetic photon spectra from 2D PIC simulations of relativistic collisionless shocks by following a sample of simulation particles in the time-varying fields  of the simulation. The low-frequency part of the spectrum scales as $\propto\omega^{2/3}$, as expected for synchrotron emission from a 2D particle distribution. Although the electromagnetic fluctuations generated by Weibel instability are on small (skin-depth) scales, the particle emission does not occur in the jitter regime. 

In retrospect, this is not surprising. The characteristic length scale of the magnetic turbulence is $\lambda\gtrsim10\comp$, and in the shock region, where most of the emission is produced, the magnetic energy reaches a fraction $\epsilon_\ditto{B}\simeq0.1$ of the upstream bulk kinetic energy. This implies $r_\ditto{L}/(\comp)=\epsilon_\ditto{B}^{-1/2}\simeq3$, where $r_\ditto{L}$ is the relativistic Larmor radius of a particle moving with the upstream flow. It follows that $K=\lambda/(r_\ditto{L}/\gamma_0)\simeq3\,\gamma_0$, so that the condition $K\ll1$ for jitter radiation is unlikely to be satisfied, even for moderately relativistic shocks. Indeed, the photon spectrum we obtain from a $\gamma_0=3$ shock is still consistent with synchrotron radiation. For electron-ion shocks, the wiggler parameter will be ${m_p/m_e}$ times larger, and we are even deeper in the synchrotron regime. Further downstream from the shock, although the magnetic field strength decays, the value of $K$ does not significantly change, since short-wavelength modes are progressively damped \citep{chang_08}, and the characteristic scale of the turbulence increases.

Although the results presented here apply to a 2D particle distribution, the main conclusions should hold for 3D configurations as well. There, we expect the low-frequency slope to be $1/3$, as in the standard synchrotron radiation. If the GRB emission results from high-energy particles accelerated in relativistic unmagnetized shocks, it seems that resorting to the jitter radiation is not a viable solution for the ``line of death'' puzzle \citep{preece_98}.

At high frequencies, our results suggest that the contribution of upstream particles to the total emission, which is usually omitted in standard models, is not negligible. It causes the radiation spectrum to be flatter than the corresponding downstream spectrum, thus partly masking the contribution of downstream thermal particles \citep{giannios_09}. This could potentially explain the absence of   clear signatures of downstream thermal emission in GRB shocks \citep{band_93}.
Simulations extending to longer times (and higher particle energies) will help to clarify these issues.
 
We remark that our calculations do not include radiative particle cooling, synchrotron self-absorption and inverse Compton radiation. Still, we have shown that the computation of synthetic spectra from self-consistent PIC simulations provides a powerful tool for studying the origin of astrophysical non-thermal emission.
\\\\
We thank J.~Arons, A.~Celotti, J.~Kirk, S.~Martins and L.~Silva for comments and suggestions, and KITP Santa Barbara for hospitality. This research was supported by NSF grants AST-
0807381 and PHY05-51164. 

\bibliography{ms}
\end{document}